\documentclass{aa}

\usepackage{txfonts}
\usepackage{graphicx}

\begin{document}

\title{Velocity asymmetries in YSO jets}
\subtitle{Intrinsic and extrinsic mechanisms}

\author{
  T. Matsakos\inst{1,2}, N. Vlahakis\inst{2}, K. Tsinganos\inst{2},
  K. Karampelas\inst{2}, C. Sauty\inst{3}, V. Cayatte\inst{3}, \\
  S. P. Matt\inst{4}, S. Massaglia\inst{5}, E. Trussoni\inst{6}, \and
  A. Mignone\inst{5}
}

\authorrunning{T. Matsakos et al.}
\titlerunning{Velocity asymmetries in YSO jets}

\institute{
  CEA, IRAMIS, Service Photons, Atomes et Mol\`ecules, F-91191 Gif-sur-Yvette,
    France \and
  IASA \& Sect. of Astrophysics, Astronomy and Mechanics, Dept. of Physics,
    University of Athens, 15784 Zografos, Athens, Greece \and
  LUTh, Observatoire de Paris, UMR 8102 du CNRS, Universit\'e Paris Diderot,
    F-92190 Meudon, France \and
  Laboratoire AIM Paris-Saclay, CEA/Irfu Universit\'e Paris-Diderot CNRS/INSU,
    91191 Gif-Sur-Yvette, France \and
  Dipartimento di Fisica, Universit\`a degli Studi di Torino, via Pietro Giuria
    1, 10125 Torino, Italy \and
  INAF/Osservatorio Astronomico di Torino, via Osservatorio 20, 10025 Pino
    Torinese, Italy
}

\date{Received ?? / Accepted ??}

\abstract{
  It is a well established fact that some YSO jets (e.g. RW Aur) display
  different propagation speeds between their blue and red shifted parts, a
  feature possibly associated with the central engine or the environment in
  which the jet propagates.
}{
  In order to understand the origin of asymmetric YSO jet velocities, we
  investigate the efficiency of two candidate mechanisms, one based on the
  intrinsic properties of the system and one based on the role of the external
  medium.
  In particular, a parallel or anti-parallel configuration between the
  protostellar magnetosphere and the disk magnetic field is considered and the
  resulting dynamics are examined both in an ideal and a resistive
  magneto-hydrodynamical (MHD) regime.
  Moreover, we explore the effects of a potential difference in the pressure of
  the environment, as a consequence of the non-uniform density distribution of
  molecular clouds.
}{
  Ideal and resistive axisymmetric numerical simulations are carried out for a
  variety of models, all of which are based on a combination of two analytical
  solutions, a disk wind and a stellar outflow.
  The initial two-component jet is modified either by inverting the orientation
  of its inner magnetic field or imposing a constant surrounding pressure.
  The velocity profiles are studied assuming steady flows as well as when strong
  time variable ejection is incorporated.
}{
  Discrepancies between the speeds of the two oppositely directed outflows can
  indeed occur both due to unaligned magnetic fields and different outer
  pressures.
  In the former case, the asymmetry appears only on the dependence of the
  velocity on the cylindrical distance, but the implied observed value is
  significantly altered when the density distribution is also taken into
  account.
  On the other hand, a non-uniform medium collimates the two jets unevenly,
  directly affecting their propagation speed.
  A further interesting feature of the pressure-confined outflow simulations is
  the formation of static knots whose spacing seems to be associated with the
  ambient pressure.
}{
  Jet velocity asymmetries are anticipated both when multipolar magnetic moments
  are present in the star-disk system as well as when non-uniform environments
  are considered.
  The latter case is an external mechanism that can easily explain the large
  time scale of the phenomenon, whereas the former one naturally relates it to
  the YSO intrinsic properties.
}


\maketitle

\section{Introduction}
  \label{sec:introduction}

Over the last few years, the two-component jet scenario emerges as a strong
candidate for describing Young Stellar Object (YSO) outflows.
Observational data of Classical T Tauri Stars (CTTS) (Edwards et al.
\cite{Edw06}; Kwan et al. \cite{Kwa07}) indicate the presence of two genres of
winds: one being ejected radially out of the central object (e.g. Sauty \&
Tsinganos \cite{Sau94}; Trussoni et al. \cite{Tru97}; Matt \& Pudritz
\cite{Mat05}) and the other being launched at a constant angle with respect to
the disk plane (e.g. Blandford \& Payne \cite{Bla82}; Tzeferacos et al.
\cite{Tze09}; Salmeron et al. \cite{Sal11}).
Consequently, CTTS outflows may be associated with either a stellar or disk
origin, or with both outflow components having comparable contributions.
In addition, such a scenario is supported by theoretical arguments (Ferreira et
al. \cite{Fer06}).
An extended disk wind is required for the explanation of the observed YSO mass
loss rates, whereas a pressure driven stellar outflow is expected to propagate
in the central region, possibly accounting for the protostellar spin down (Matt
\& Pudritz \cite{Mat08}; Sauty et al. \cite{Sau11}).
Numerical simulations have been also employed recently to investigate the
various aspects of two-component jets (Meliani et al. \cite{Mel06}; Fendt
\cite{Fen09}; Matsakos et al. \cite{Tit09}, hereafter M09).

Detailed observations of some YSO with bipolar flows have shown peculiar
velocity asymmetries between the blue and red shifted regions (e.g. Woitas et
al. \cite{Woi02}; Coffey et al. \cite{Cof04}; Perrin et al. \cite{Per07}).
In particular, recent estimates of the RW Aur jet indicate that the speed of the
approaching lobe is roughly 50\% higher as compared to the receding one
(Hartigan \& Hillenbrand \cite{Har09}).
Although the above studies suggest that the asymmetry originates close to the
source, observations of the same object carried out by Melnikov et al.
(\cite{Mel09}) point out a similar mass outflow rate for the two opposite jets,
concluding that their different speeds are due to environmental effects.
Thus, it is still an open question whether the discrepancy between the two
hemispheres relies on an intrinsic property, such as the magnetic field
configuration, or an external factor, such as the physical conditions of the
surrounding medium.

Spectropolarimetric measurements of T Tauri stars suggest multipolar
magnetospheres, with the dipolar component not always parallel to the rotation
axis (Valenti \& Johns-Krull \cite{Val04}; Daou et al. \cite{Dao06}; Yang et al.
\cite{Yan07}; Donati \& Landstreet \cite{Don09}).
Non-equatorially symmetric field topologies may affect both the way that matter
accretes on the protostar (Long et al. \cite{Lon07}; Long et al. \cite{Lon08})
and the jet launching mechanisms.
For jets, it is not clear whether the magnetic field asymmetry persists for
timescales comparable to the jet propagation timescales (decades-centuries).
In fact, a periodic stellar activity could be directly associated with jet
variability.
In a similar context, Lovelace et al. (\cite{Lov10}) simulated the central
region of YSOs assuming complex magnetic fields.
In one interesting case, where the magnetosphere is a combination of dipolar and
quadrupolar moments, they find the extreme scenario of one-sided conical
outflows being ejected from the star-disk interaction interface.
On the other hand, the accretion disk could also have a quadrupolar magnetic
field originating either from the dynamo mechanism or from advection (Aburihan
et al. \cite{Abu01}).
In fact, it has been shown that non-bipolar disk field topologies can be a
potential source for asymmetry in AGN jets (Wang et al. \cite{Wan92};
Chagelishvili et al. \cite{Cha96}).

The external medium may play a relevant role in the dynamical propagation of the
outflow; we refer again to the detailed HST observations of RW Aur by Melnikov
et al. (\cite{Mel09}), whose findings can be consistent with the effects of
inhomogeneities in the environmental conditions.  
Clumpy and filamentary structures are observed in molecular clouds over a wide
range of scalelenghts, implying the presence of an external medium that
surrounds the jets.
In addition, an environment affecting jet propagation may be present but remain
almost undetectable in observations, as discussed recently by Te\c{s}ileanu et
al. (\cite{Tes12}). 
Theoretical arguments suggest that density anisotropies in the vicinity of YSOs
could collimate YSO outflows and even induce oscillations in the jet's cross
section (K\"onigl \cite{Kon82}).
Numerical simulations have investigated the effects of the surrounding gas, such
as a collapsing environment whose ram pressure collimates a spherical wind
(Delamarter et al. \cite{Del00}) or an isothermal medium with a toroidal density
distribution that results in a cylindrically shaped outflow (Frank \&
Noriega-Crespo \cite{Fra94}; Frank \& Mellema \cite{Fra96}).
Another class of simulations has studied the collimating role of a vertical
outer magnetic field, whose magnetic pressure effectively confines an isotropic
stellar wind (Matt \& B\"ohm \cite{Ma03a}; Matt et al. \cite{Ma03b}).
We note that although the external pressure is not thought to be responsible for
the jet collimation observed at larger scales, it could still affect it and
hence modify the propagation speed.

The goal of the present work is to study the feature of the velocity asymmetry
within the context of the two-component jet models presented in M09.
We take advantage of both theoretical and numerical approaches (Gracia et al.
\cite{Gra06}), setting as initial conditions a combination of two analytical YSO
outflow solutions.
Tabulated data of such MHD jets have been derived using the self-similarity
assumption (Vlahakis \& Tsinganos \cite{Vla98}) and are available for both disk
winds and stellar jets.
We employ the Analytical Disk Outflow (ADO) solution from Vlahakis et al.
(\cite{Vla00}) and the Analytical Stellar Outflow (ASO) model from Sauty et al.
(\cite{Sau02}).

Matsakos et al. (\cite{Tit08}) have addressed the topological stability, as well
as several physical and numerical properties of each class of solutions
separately.
In M09 the two complementary outflows were properly mixed inside the
computational box, such that the stellar outflow dominated the inner regions and
the disk wind the outer.
The stability and co-evolution of several dual component jet cases was
investigated as a function of the mixing parameters and the enforced time
variability.
Furthermore, the introduction of flow fluctuations generated shocks, whose large
scale structure demonstrated a strong resemblance to real YSO jet knots.

In this paper, two main scenarios are examined as possible mechanisms to produce
velocity asymmetries.
The first one refers to intrinsic YSO properties introducing distinct magnetic
field topologies at each side of the equator.
The second one refers to external effects arising from differences in the
ambient pressure.

\begin{figure*}
  \resizebox{\hsize}{!}{\includegraphics{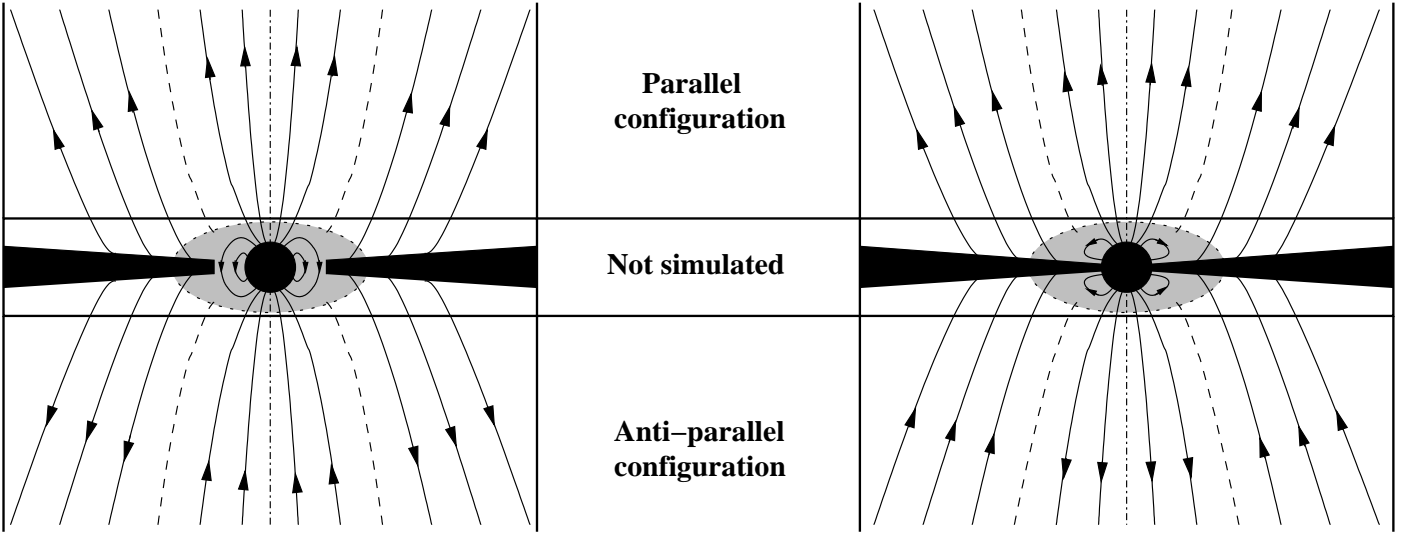}}
  \caption{Left panel: a quadrupolar disk field and a dipolar stellar
    magnetosphere.
    Right panel: a bipolar disk field and a quadrupolar stellar magnetosphere.
    The northern hemisphere of either case is referred to as a parallel
    configuration, whereas the southern as an anti-parallel.
    The dashed line indicates the surface that separates the disk wind from the
    stellar outflow.
    The simulations do not include the central region of the YSO, and hence we
    do not attempt to describe the star-disk interaction that takes place inside
    the grey ellipse.
    In addition, the requirement of $\nabla\cdot\vec B = 0$ means that field
    lines must return along the accretion disk.}
  \label{fig:parallel_antiparallel}
\end{figure*}
In the first case, presented in Fig.~\ref{fig:parallel_antiparallel}, each
hemisphere is considered to have one of the following two magnetic field
configurations: the field lines coming out of the stellar surface are parallel
(north) or anti-parallel (south) to the large scale magnetic field.
This implies a quadrupolar disk field and a dipolar magnetosphere (left), or
equivalently a bipolar disk field surrounding a magnetic quadrupole (right).

\begin{figure}
  \resizebox{\hsize}{!}{\includegraphics{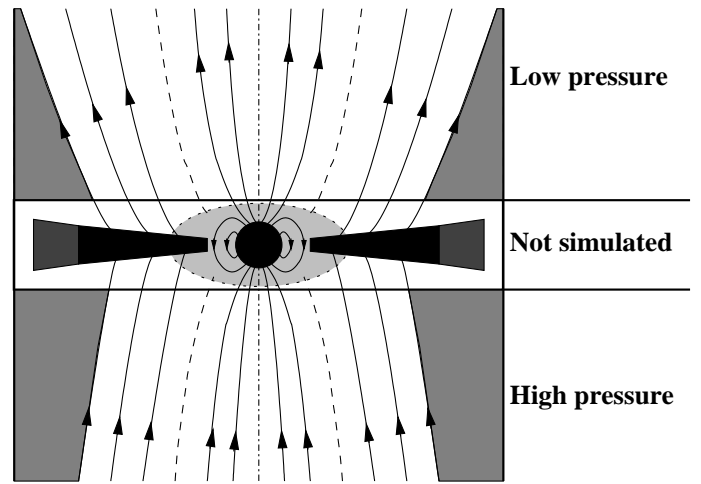}}
  \caption{The northern hemisphere assumes a lower pressure, and hence a lower
    degree of collimation, as compared to the southern.
    The sketch is not in scale, the effects of this mechanism will hold at any
    distance wherein distinct environments are encountered.
    The magnetic field topology here is simple, a dipolar stellar magnetosphere
    is surrounded by a bipolar disk field.}
  \label{fig:external_pressure}
\end{figure}
In the second scenario, shown in Fig.~\ref{fig:external_pressure}, the
properties of the medium along the propagation axis are assumed to affect the
collimation of the jet.
Specifically, the outer pressure could modify the diameter of the outflow and in
turn it would determine the wind speed.
This mechanism does not depend on any YSO time scale, and it could appear at any
height that the jet encounters heterogeneous environment.
We note that although a thermal pressure of a hot tenuous medium is used in the
simulations, this could equivalently represent the pressure due to a cold dense
environment, turbulence or even a large scale magnetic field.
Stute et al. (\cite{Stu08}) truncated ADO solutions imposing a weaker outflow of
the same type at outer radii.
Here, we extend their study assuming a constant surrounding pressure instead,
also with different values at each jet direction.

All our models focus on large enough scales avoiding the complicated dynamics of
the star-disk interaction (e.g. Bessolaz et al. \cite{Bes08}; Zanni et al.
\cite{Zan09}; Lovelace et al. \cite{Lov10}).
Essentially, these mechanisms are included merely as boundary conditions, an
approach frequently adopted in the literature of jet propagation studies (e.g.
Ouyed et al. \cite{Ouy03}; Fendt \cite{Fen06}; Te\c{s}ileanu et al.
\cite{Tes09}).

Simulations are carried out to study the time evolution as well as the potential
final steady states, comparing the results obtained between the two sides of the
equatorial plane.
We are especially looking for emerging asymmetries in the vertical velocity
profiles.
The resistive MHD regime is of particular interest (\v{C}emelji\'c et al.
\cite{Cem08}) since reconnection plays a key role, mainly at the locations where
the magnetic field inverts sign.
Furthermore, all cases are re-examined assuming that the flow has a strong time
variable character, an effect that produces discontinuities along its axis.
Note that we do not attempt to model the observed properties of the RW Aur jet,
but rather to investigate in a general manner the intrinsic or extrinsic
physical processes that could underlie the velocity asymmetry phenomenon.

The paper is structured as follows.
Section \S\ref{sec:theory} describes the jet models and provides information on
the numerical setup.
Section \S\ref{sec:results} presents and discusses the obtained results while
section \S\ref{sec:conclusions} reports the conclusions of this work.

\section{Numerical models and setup}
  \label{sec:theory}

\subsection{MHD equations}
  \label{sec:mhd}

The non-dimensional resistive MHD equations written for the primitive variables
are:
\begin{equation}
  \frac{\partial\rho}{\partial t} + \nabla \cdot (\rho \vec V) = 0\,,
\end{equation}
\begin{equation}
  \frac{\partial\vec V}{\partial t} + (\vec V \cdot \nabla)\vec V
    + \frac{1}{\rho}\vec B \times\vec J
    + \frac{1}{\rho}\nabla P = - \nabla \Phi\,,
\end{equation}
\begin{equation}
  \frac{\partial P}{\partial t} + \vec V \cdot \nabla P
    + \Gamma P \nabla \cdot \vec V = (\Gamma - 1)\eta\vec J^2 + \Lambda\,,
\end{equation}
\begin{equation}
  \frac{\partial\vec B}{\partial t} + \nabla \times (\vec B \times \vec V)
    = -\eta\nabla\times\vec J\,.
\end{equation}
The quantities $\rho$, $P$ and $\vec V$ are the density, pressure and velocity
respectively.
The magnetic field $\vec B$, which includes the factor $(4\pi)^{-1/2}$,
satisfies the condition $\nabla\cdot\vec B = 0$.
Moreover, the electric current is defined as $\vec J = \nabla\times\vec B$, and
has also absorbed a factor $(4\pi)^{-1/2}$.
The gravitational potential is given by $\Phi = -GM/R$, where $G$ is the
gravitational constant, $M$ the mass of the central object and $R$ the spherical
radius.
The magnetic resistivity tensor has been assumed diagonal, with all its elements
equal to the scalar quantity $\eta$.
Finally, $\Lambda$ represents the volumetric energy gain/loss terms and
$\Gamma = 5/3$ is the ratio of the specific heats.

We adopt the axisymmetric cylindrical coordinates $(r,z)$ and we will express
all quantities in code units.
A code variable $U$ is converted to its physical units $U'$ with the help of
$U_0$, i.e. $U' = U_0U$.
The constants $U_0$ are specified such that the values of the quantities in the
final configuration of the simulations match the typical values of real jets,
such as observed velocities and densities.
Specifically, $r_0 = z_0 = 1.0\,\mathrm{AU}$,
$\rho_0 = 2.1\cdot10^{-18}\,\mathrm{g}\,\mathrm{cm}^{-3}$,
$V_0 = 4.0\,\mathrm{km}\,\mathrm{s}^{-1}$,
$P_0 = 3.3\cdot10^{-7}\,\mathrm{dyne}\,\mathrm{cm}^{-2}$,
$B_0 = 2.0\cdot10^{-3}\,\mathrm{G}$ and $t_0 = 1.2\,\mathrm{yr}$.

\subsection{Analytical solutions}

The employed ADO solution, denoted with subscript D, is appropriate for
describing disk winds.
It is a steady state radially self-similar MHD outflow, namely, each quantity
has a certain scaling along the disk radius based on the Keplerian rotation
profile.
Therefore, if all flow variables are known along a particular fieldline the
spatial distribution of the entire solution can be reconstructed.
This type of models have conical critical surfaces and consequently diverge on
the axis.
However, we substitute the inner region with the ASO solution, denoted with S, a
meridionally self-similar jet.
In this case, the critical surfaces have a spherical shape that effectively
models stellar outflows.
Some kind of energy input is required at the base of the wind to accelerate the
plasma, but we do not treat this region because we simulate high altitudes that
correspond to its super Alfv\'enic domain.
Since the ADO solution is described by the polytropic index $\Gamma = 1.05$, we
specify the same value in the simulations which implies an energy source term
$\Lambda = (5/3 - \Gamma)P\nabla\cdot\vec V$.
We have verified that this does not significantly affect the dynamics (see also
Matsakos et al. \cite{Tit08}).
In other words, although thermal effects are thought to be crucial for the mass
loading close to the disk surface (Ferreira \& Casse \cite{Fer04}), we can
neglect them above it where the magneto-centrifugal mechanism takes over.
The adopted analytical solutions together with their parameters are presented in
detail in M09.

\subsection{Normalization and mixing}

The two-component jet model parameters can be classified into two categories.
The first one describes the relative normalization of the analytical solutions
with the help of the following ratios:
\begin{equation}
  \ell_\mathrm{L} = \frac{R_*}{r_*}\,,\quad
  \ell_\mathrm{V} = \frac{V_{\mathrm{S}*}}{V_{\mathrm{D}*}}\,,\quad
  \ell_\mathrm{B} = \frac{B_{\mathrm{S}*}}{B_{\mathrm{D}*}}\,,
\end{equation}
where $R_*$ is the Alfv\'enic spherical radius of the ASO model, $r_*$ the
cylindrical radius of the Alfv\'enic surface on a specific fieldline of the ADO
solution and the star denotes the characteristic values at these locations.
The subscripts L, V and B stand for length, velocity and magnetic field,
respectively.
We specify $\ell_\mathrm{V} = 5.96$, such that the two solutions correspond to
the same protostellar gravitational field.
Following the guidelines of M09 we set $\ell_\mathrm{L} = 0.1$.
Note that due to the non-existence of a characteristic length scale in the ADO
solution, the value of this parameter does not play any role in the combination
of the two wind components.
We also define $\ell_\mathrm{B} = 0.01$, in order to provide comparable magnetic
fields for the two outflows in the initial conditions of the numerical setup.
Larger or lower values for this parameter would result in either a dominant
stellar outflow and a negligible disk wind or the opposite, respectively.

The second class of parameters controls the combination of the two analytical
models.
In order to achieve a smoothly varying magnetic field and ensure the dominance
of the protostellar magnetosphere in the central region, we follow an improved
mixing process as compared to the one presented in M09.
The merging depends on the magnetic flux functions $A_\mathrm{D}$ and
$A_\mathrm{S}$, which label the field lines of each initial one-component
solution.
Once a total magnetic flux $A$ is defined, then the poloidal component of the
magnetic field is given by:
\begin{equation}
  \vec{B}_\mathrm{p} = (\nabla A\times\hat\phi)/r\,,
  \label{eq:Bp}
\end{equation}
where $\hat\phi$ is the unit toroidal vector.
However, a steep transition from one profile to the other could create
artificial gradients in the flux which would be reflected in the magnetic field.
As a result, $\vec B$ could end up being excessively strong or even inverted
across the matching surface.

Therefore, the footpoint of the fieldline that marks the transition,
$A_\mathrm{mix}$, is chosen at a location where both $A_\mathrm{D}$ and
$A_\mathrm{S}$ have roughly the same slope.
Then, a first approximation of the total magnetic flux is defined:
$A_1 = A_\mathrm{D} + A_\mathrm{S} + A_\mathrm{C}$, where $A_\mathrm{C}$ is a
constant that helps normalize $A_\mathrm{D}$ and $A_\mathrm{S}$.
We require that the disk and stellar fields are exponentially damped in the
central and outer regions respectively, which leads to an improved approximation
for the magnetic flux:
\begin{equation}
  A_2 = \left\{1 - \exp
    \left[-\left(\frac{A_1}{A_\mathrm{mix}}\right)^2\right]\right\}
    A_\mathrm{D} + \exp
    \left[-\left(\frac{A_1}{A_\mathrm{mix}}\right)^2\right]
    A_\mathrm{S}\,.
\end{equation}
Finally, with the help of $A_2$ all two-component physical variables, $U$, are
initialized using the following gaussian-type mixing function:
\begin{equation}
  U = \left\{1-\exp
    \left[-\left(\frac{A_2}{A_\mathrm{mix}}\right)^2\right]\right\}
    U_\mathrm{D} + \exp
    \left[-\left(\frac{A_2}{A_\mathrm{mix}}\right)^2 \right]
    U_\mathrm{S}\,.
\end{equation}
This relation is also applied to generate the total initial two-component
magnetic flux, $A$.
The poloidal field is then derived from Eq.~(\ref{eq:Bp}) and it is divergence
free by definition.

Moreover, since accretion and protostellar variability are expected to introduce
fluctuations in the ejection, the velocity prescribed on the bottom boundary is
multiplied with the following function:
\begin{equation}
  f_\mathrm{S}(r, t) =
    1 + p\sin\left(\frac{2\pi t}{T_\mathrm{var}}\right)
    \exp\left[-\left(\frac{r}{r_\mathrm{var}}\right)^2\right]\,,
  \label{eq:variability}
\end{equation}
where $T_\mathrm{var} = 2.5$ is the period of the pulsation and
$r_\mathrm{var} = 5$ is roughly the cylindrical radius at which the matching
surface intersects the lower boundary of the computational box.
The fractional variability is set to $p = 0.5$ in the time variable models, in
order to check the stability in the presence of strong perturbations and enhance
reconnection when resistivity is included.
Essentially, Eq.~(\ref{eq:variability}) provides a gaussian spatial distribution
of a periodic velocity variability of $\sim$$3\,\mathrm{yr}$, that produces
shocks and knot-like structures.

\subsection{Numerical models}
  \label{sec:models}

The anti-parallel configuration is initialized assuming a sudden inversion of
the magnetic field beyond the fieldline rooted at
$r_\mathrm{inv} = 5$.
In non-resistive MHD the two systems are entirely equivalent due to the
invariance of the axisymmetric MHD equations under the following
transformations:
i) $\vec{B}\to-\vec{B}$, ii) $B_\phi\to-B_\phi$ and $V_\phi\to-V_\phi$.
The finite values of the electric conductivity and its dependence on quantities
such as $\vec{J}$ is an unknown factor in YSO outflows.
Therefore, we follow a simple approach applying a constant physical resistivity,
with the value $\eta = 1.0$, an approximation adequate to provide a rough
estimate of how the system evolves in the presence of non-ideal effects.
As explained in \S\ref{sec:results}, this value can be considered small because
it ensures the dominance of the advection terms over the resistive ones by
several orders of magnitude throughout most parts of the computational box.
However, this does not hold true on reconnection separatrices and hence its
impact cannot be neglected there.
Finally, we note that even in the ideal MHD cases, the finite size of the grid
cells inevitably introduces artificial reconnection.
We minimize its effects by increasing the resolution and we estimate its
influence by comparing results of different mesh refinements.

On the other hand, the truncation of the jet for the low pressure hemisphere is
initialized with the following process.
We calculate the total pressure of the outflow, thermal plus magnetic, at the
location $r_\mathrm{P} = 30$ on the bottom boundary\footnote{
  The plasma-$\beta$ (thermal over magnetic pressure) is approximately equal to
  $0.03$ there.
}.
Then, a constant pressure $P_\mathrm{L}$ of the same magnitude is applied on the
domain beyond the limiting fieldline rooted at $r_\mathrm{P}$.
In addition, the density is arbitrarily reduced by two orders of magnitude,
whereas the velocity and magnetic fields are set to zero.
Although pressure equilibrium holds at the bottom boundary, the weaker field and
jet pressure at higher altitudes cannot compensate for the push of the external
medium.
As a result, a radial collapse is expected during the first steps of the
simulation until the horizontal forces are balanced.
For the jet propagating in the opposite direction, we assume another simulation
case where the truncation is still at $r_\mathrm{P}$, but a higher outer
pressure is applied with the value $P_\mathrm{H} = 2P_\mathrm{L}$.
Finally, we include a case where the medium has $P_\mathrm{VH} = 6P_\mathrm{L}$
in order to discuss the morphology of the final steady state configuration.

\begin{table}
  \caption{The numerical models, the name of which follows from the initials of
    their description.}
  \label{tab:models}
  \centering 
  \begin{tabular}{lcccc}
    \hline
    \hline
    Name & Variability  & MHD        & $B$ orientation  & Outer $P$ \\
    \hline
    NIP   & Non-variable & Ideal      & Parallel        & - \\
    NIA   & Non-variable & Ideal      & Anti-parallel   & - \\
    NRP   & Non-variable & Resistive  & Parallel        & - \\
    NRA   & Non-variable & Resistive  & Anti-parallel   & - \\
    VIP   & Variable     & Ideal      & Parallel        & - \\
    VIA   & Variable     & Ideal      & Anti-parallel   & - \\
    VRP   & Variable     & Resistive  & Parallel        & - \\
    VRA   & Variable     & Resistive  & Anti-parallel   & - \\
    \hline
    NIPL  & Non-variable & Ideal      & Parallel        & Low \\
    NIPH  & Non-variable & Ideal      & Parallel        & High \\
    NIPVH & Non-variable & Ideal      & Parallel        & Very High \\
    VIPL  & Variable     & Ideal      & Parallel        & Low \\
    VIPH  & Variable     & Ideal      & Parallel        & High \\
    VIPVH & Variable     & Ideal      & Parallel        & Very High \\
    \hline
  \end{tabular}
\end{table}
Table~\ref{tab:models} lists the fourteen two-component jet models considered,
along with a brief description of their setup.
In particular, models NIP and NIA focus on the effects of the parallel and
anti-parallel magnetic field configuration in the ideal MHD formulation.
Cases NRP and NRA include resistivity in order to understand the role of
magnetic reconnection and Ohmic heating, especially in the regions where the
magnetic field inverts sign.
On the other hand, different values of an external pressure are imposed in
models NIPL, NIPH and VIPVH to investigate the effects of distinct environments.
We also consider the time-variable version of all previous cases to examine
their behavior in the presence of flow fluctuations, i.e. models VIP, VIA, VRP,
VRA, VIPL, VIPH and VIPVH.

\subsection{Numerical setup}
  \label{sec:setup}

The simulations are performed with PLUTO\footnote{Freely available at
\texttt{http://plutocode.ph.unito.it}}, a modular shock-capturing numerical code
for computational astrophysics (Mignone et al. \cite{Mig07}; Mignone et al.
\cite{Mig12}).
Cylindrical coordinates are chosen in 2.5D and second order accuracy is
specified in both time and space.
All simulations adopt the TVDLF solver (Lax-Friedrichs scheme), a choice that is
essential to retain stability but comes at the cost of increased numerical
diffusivity.
Outflow conditions are specified at the top and right boundaries, axisymmetric
at the left, and all variables are kept fixed to their initial values on the
bottom boundary.
The size of our domain is $0 \le r \le 100$ in the radial direction and
$10 \le z \le 310$ in the vertical.
We are interested in radial distances less that $50$, but a larger box ensures
that any boundary effects coming from the imposed zero derivative of $\vec B$
will not influence our results (Zanni et al. \cite{Zan07}).
Note that the detailed launching mechanisms of each component are not taken into
account since the computational box starts from a finite height.
The grid resolution is set to $500 \times 1500$ and the simulations are carried
out until $t_\mathrm{stop} = 30$.
We point out that the non-variable and non-pressure-confined simulations reach
an exact steady state before $t = 10$.
In M09, we have studied the steady state of such two-component jet models in
larger time scales and we have found that the configuration is time-independent
to very high accuracy.
The pressure-confined models require an order of magnitude larger evolution time
to attain a quasi steady outcome, and hence we specify $t_\mathrm{stop} = 300$
and a resolution of $300\times900$ for efficiency purposes.

\section{Results}
  \label{sec:results}

As an overall remark, all non-truncated models reach steady states in an
equivalent physical time of a few years.
The final configuration of the disk wind remains close to the initial setup,
demonstrating the stability of the ADO solution despite its significant
modification.
On the other hand, the hot stellar component gets compressed around the axis,
which also results in an increase of the flow speed there.
Even in the cases where the flow is strongly fluctuating, the two-component jet
retains on average the properties of the unperturbed model.
A steady shock that forms along a somewhat diagonal line at the lower part of
the box is found to causally disconnect the jet launching regions from the
propagation domain.
Although a different mixing has been assumed, all the above features are common
to the two-component jet models presented in M09, wherein a detailed discussion
can be found.

\subsection{Parallel/anti-parallel configurations in ideal MHD}

\begin{figure}
  \resizebox{\hsize}{!}{\includegraphics{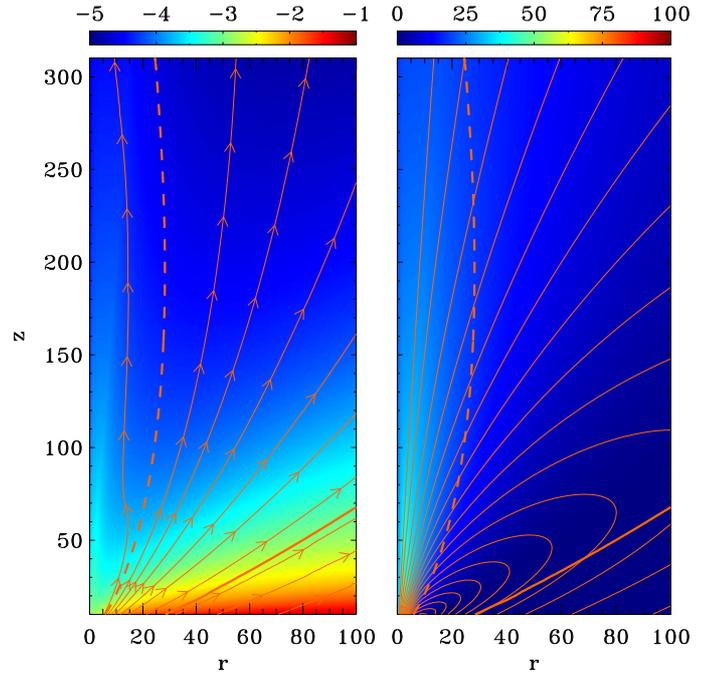}}
  \caption{Logarithmic density contours along with magnetic field lines (left)
    and vertical velocity contours together with poloidal current lines (right)
    for the initial conditions of model NIP.
    The dashed line indicates the surface $A_\mathrm{mix}$ around which the
    mixing is assumed, and also marks the polarity reversal for the
    anti-parallel configurations.
    The thick solid line shows the separatrix beyond which a constant pressure
    is initially applied on the pressure-confined models.
    The initial setup of NIA is identical except for the opposite sign of
    $\vec{B}$ and $\vec{J}$ in the central regions.}
  \label{fig:NIP_initial}
\end{figure}
The left panel of Fig.~\ref{fig:NIP_initial} presents the density and the
magnetic field lines of the starting setup of NIP, whereas the right panel shows
the distribution of the vertical velocity and the poloidal current contours.
The initial configuration of NIA has an identical morphology apart from the
reversal of $\vec B$ and $\vec J$ that takes place inside the region marked with
a dashed fieldline.
Essentially, this surface separates the hot stellar outflow from the cold
magneto-centrifugal disk wind, that is supposed to carry most of the mass and
angular momentum extracted from the system.
Moreover, the pressure driven flow is faster as compared to the outer parts of
the two-component model, a property associated with the normalization and the
acceleration efficiency of the specific analytical solutions employed.
In relation to the discussion on the mixing function, Fig.~\ref{fig:NIP_initial}
demonstrates that there is a restricted choice for the location of the matching
surface due to the requirement that both solutions ought to have there a
magnetic field of a similar shape and magnitude.
In other words, the roughly vertical field of the stellar component cannot be
matched smoothly with the bent lines coming out of the disk at outer radii.
Finally, the thick solid fieldline represents the truncation surface that is
imposed in the simulations which study the effects of an external pressure.

\begin{figure}
  \resizebox{\hsize}{!}{\includegraphics{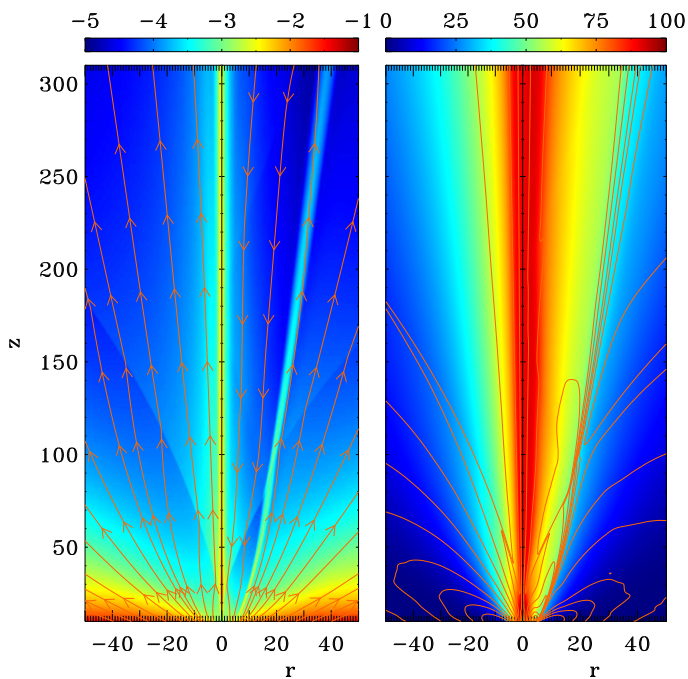}}
  \caption{In the left panel, the image shows the logarithmic density and the
    magnetic field lines for the final steady states reached for models NIP
    (left hand side; parallel configuration) and NIA (right hand side;
    anti-parallel configuration).
    In the right panel, the vertical velocity distribution and electric current
    contours are shown for the same models.
    Each panel displays side by side both hemispheres of the YSO system,
    described in Fig.~\ref{fig:parallel_antiparallel}, in order to compare the
    steady states reached by the two models.
    Matter accumulates along the current sheet of NIA reducing the density of
    inner radii as well.
    In addition, its velocity magnitude is similar to NIP although with a more
    extended radial profile.}
  \label{fig:NIP-NIA_final}
\end{figure}
Each one of the two panels of Fig.~\ref{fig:NIP-NIA_final} shows side by side
the jet evolution of the parallel and anti-parallel configuration described in
Fig.~\ref{fig:parallel_antiparallel}.
In particular, the left pair shows the density and magnetic field lines of the
final steady states reached by NIP (left) and NIA (right).
The flipping of the field is evident in the anti-parallel case, although the
general structure is maintained.
From the theoretical point of view of ideal MHD, the magnetic field reversal
($\vec{B}\to-\vec{B}$) of a flux tube would not change the dynamics.
However, the thickness of the current sheet that would form on its surface
cannot be assumed zero in a physical situation.

The simulation of NIA gives a decreased value of density at inner radii and a
peak along the field inversion surface.
The latter feature forms simultaneously at all heights during the first steps of
the simulation, and it appears more prominent in lower resolutions.
In fact, numerical reconnection inevitably appears when a curved current sheet
is dragged through a square cell grid.
The field, and in particular the strong toroidal component that dominates the
magnetic pressure, are destroyed at the interface giving rise to a perpendicular
force.
Matter accumulates along that surface and eventually the thermal pressure
compensates the lack of the magnetic one.
More accurate solvers and higher resolution would treat in a more consistent way
the separatrix of the inversion of $\vec{B}$.
However, on the one hand the applied physical resistivity of the simulations
presented below is well above the numerical diffusivity, and on the other,
``smoothing out'' effects are an important stability factor for the initial
two-component magnetic field.

The right pair of Fig.~\ref{fig:NIP-NIA_final} plots the final vertical velocity
distribution and the poloidal currents for the corresponding models shown on the
left.
A wider radial profile of $V_z$ is observed in NIA as compared to NIP.
Additionally, apart from the current sheet that forms in both models along the
weak diagonal shock, another one also appears on the surface of the magnetic
field reversal, as expected according to the previous discussion.

\subsection{Parallel/anti-parallel configurations in resistive MHD}

Based on the analysis of \v{C}emelji\'c et al. (\cite{Cem08}) we consider the
magnetic Reynolds number $R_\mathrm{m} = rV_z/\eta$ and the quantity
$R_\beta = (P/B^2)R_\mathrm{m}$, which measure the contribution of the resistive
terms in the induction and energy equations, respectively.

\begin{figure}
  \resizebox{\hsize}{!}{\includegraphics{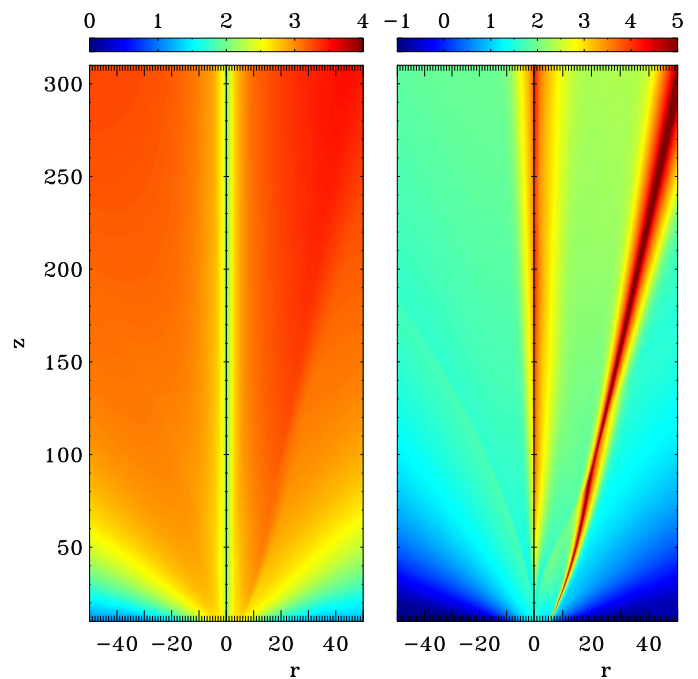}}
  \caption{The logarithmic distribution of the magnetic Reynolds number
    $R_\mathrm{m} = rV_z/\eta$ (left panel) and $R_\beta = (P/B^2)R_\mathrm{m}$
    (right panel) for both YSO hemispheres, i.e. NRP (on the left of each pair)
    and NRA (on the right of each pair).
    Note that their definition does not include the length scale of the magnetic
    field reversal, but we can still observe that the resistive effects are
    small over most of the computational domain.}
  \label{fig:NRP-NRA_Rm-Rb}
\end{figure}
Fig.~\ref{fig:NRP-NRA_Rm-Rb} displays $\log(R_\mathrm{m})$ (left pair) and
$\log(R_\beta)$ (right pair) for the final states of NRP (left) and NRA (right).
Clearly, the magnetic Reynolds number is large in most parts of the
computational domain, i.e. $R_\mathrm{m} \gg 1$, indicating that the diffusion
of the magnetic field is not of particular importance.
However, note that this definition of $R_\mathrm{m}$ does not take into account
the magnetic field reversal whose characteristic length is much smaller than
$r$ and hence the above argument excludes this region.
Therefore, approximating resistivity with a small constant value is a useful
approach to allow reconnection in our models without significantly affecting the
validity of the ideal MHD results of the inner and outer regions.

Moreover, we observe that the condition $10 < R_\beta < 100$ holds in most parts
of our domain indicating that Ohmic heating is not important either.
The value of $R_\beta$ is also ill-defined in the anti-parallel case, since it
is based on $R_\mathrm{m}$.
Nevertheless, note that $R_\beta$ is large at the axis where the plasma is hot
and along the matching surface that the magnetic field reconnects, due to the
high values of the plasma-$\beta$ there.

\begin{figure}
  \resizebox{\hsize}{!}{\includegraphics{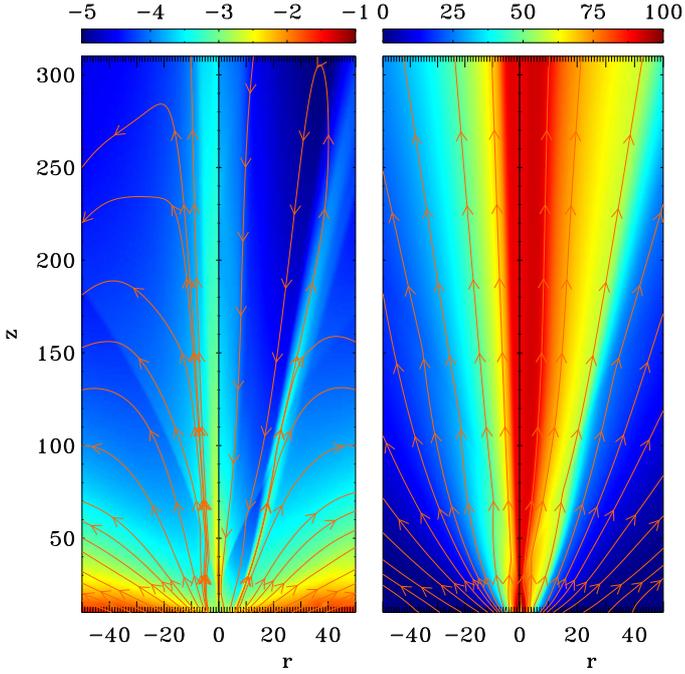}}
  \caption{Side by side logarithmic density contours and magnetic field lines
    (left panel) for the two opposite propagating jets, i.e. NRP (on the left)
    and NRA (on the right).
    Contours of $V_z$ and streamlines (right panel) for the same models.
    Model NRA displays higher vertical velocity around the transition region as
    compared to NRP.}
  \label{fig:NRP-NRA_final}
\end{figure}
Fig.~\ref{fig:NRP-NRA_final} shows the distribution of $\log(\rho)$ and the
magnetic field lines (left pair) of the final steady states of NRP (left) and
NRA (right).
The right pair displays the streamlines over the contours of $V_z$ for the same
models.
Initially, the asymmetric magnetic topologies lead to distinct current
distributions.
Reconnection is manifested at the regions where the magnetic field inverts sign,
driving a different temporal evolution between the two hemispheres.
Evidently, the presence of resistive effects in the anti-parallel case results
in a spatial readjustment of the velocity profile in the inner regions.

Another observed feature is the reversal of $B_z$ that takes place at the upper
and outer domains of both NRP and NRA.
This is a steady state configuration that originates from the resistive effects
at the base of the outflow and a shear in the velocity profile of the flow.
In particular, although the two vector fields, $\vec B$ and $\vec V$, are
parallel at the bottom boundary, the low $R_\mathrm{m}$ values suggests that
diffusion dominates advection at the zones right above it.
As a consequence, a small angle forms between the poloidal components of
$\vec B$ and $\vec V$ before the magnetic Reynolds number of the flow acquires
high values.
Shearing effects increase this angle resulting in a negative $B_z$ component.

\begin{figure}
  \resizebox{\hsize}{!}{\includegraphics{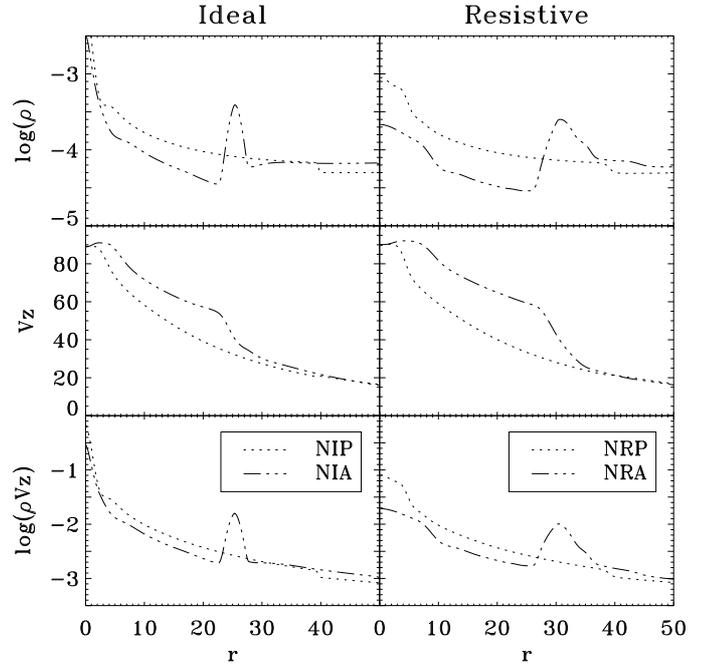}}
  \caption{Radial profile of $\log(\rho)$, $V_z$ and $\log(\rho V_z)$ at
    $z = 160$ for models NIP (left; dotted line), NIA (left; dashed dotted), NRP
    (right; dotted) and NRA (right; dashed dotted).
    Resistive effects are negligible for the dynamics of the parallel case,
    while they amplify the features observed in the ideal anti-parallel one.}
  \label{fig:radial_profile}
\end{figure}
Fig.~\ref{fig:radial_profile} plots $V_z$ (middle) and the logarithms of $\rho$
(top), and $\rho V_z$ (bottom) along the radial direction at $z = 160$ for
models NIP (left; dotted line), NIA (left; dashed dotted line), NRP (right;
dotted line) and NRA (right; dashed dotted line).
From the parallel setups we deduce that the prescribed value of resistivity has
a minor effect on dynamics.
On the contrary, non-ideal MHD effects amplify the features of the anti-parallel
configuration, namely the decrease of the density at inner radii accompanied by
an increase of the speed.
In addition, the peak observed in $\rho$ along the current sheet is more
extended in NRA due to the physical reconnection applied.
The mass flux profiles also reflect the asymmetry, despite the fact that $\rho$
and $V_z$ change in opposite directions.
Notably, beyond the matching surface all four models demonstrate similar
behaviors.

\subsection{Time variable flows in parallel/anti-parallel configurations}

This section investigates the previously presented models when a time variable
velocity is applied on the bottom boundary, localized at the stellar component.
The simulations are performed until the final time of the corresponding
unperturbed cases is reached, during which several shocks propagate throughout
the computational domain.

\begin{figure*}
  \resizebox{\hsize}{!}{\includegraphics{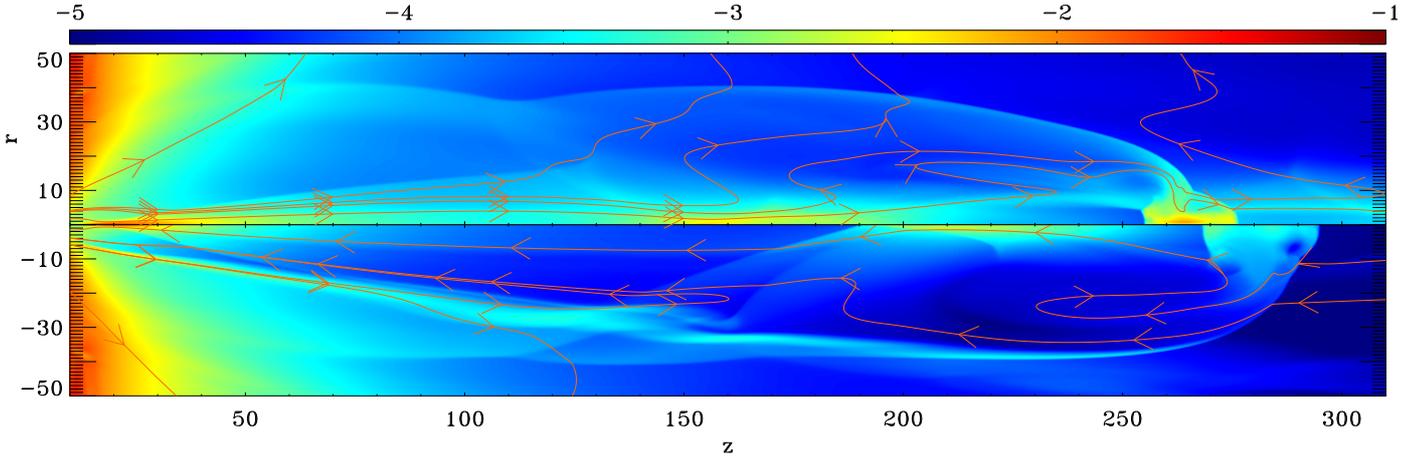}}
  \caption{Logarithmic density and field lines for models VRP (top) and VRA
    (bottom).
    The two jets correspond to the two hemispheres of a YSO and are shown side
    by side for comparison.
    The shock front and matter condensations along the axis are found at
    different locations between the two cases.
    The stellar component of VRA displays a lower inner density as compared to
    VRP, whereas the inner part of its disk wind has higher values of $\rho$ as
    matter accumulated along the current sheet is blown outwards from radial
    shocks.}
  \label{fig:VRP-VRA_rho-B}
\end{figure*}
Fig.~\ref{fig:VRP-VRA_rho-B} compares the logarithmic density contours and the
magnetic field between models VRP (top) and VRA (bottom).
The shock front is slightly faster in the anti-parallel configuration and seems
to have a larger radius too.
High matter concentrations along the jet axis are observed at substantially
different heights, i.e. $z = 160$ for VRP and $z = 210$ for VRA.
Furthermore, higher density is also found beyond the matching fieldline of VRA
suggesting that the disk wind is affected as well.
In fact, the matter trapped along the current sheet is pushed radially out by
the inner shocks that travel across the jet flow.

\begin{figure}
  \resizebox{\hsize}{!}{\includegraphics{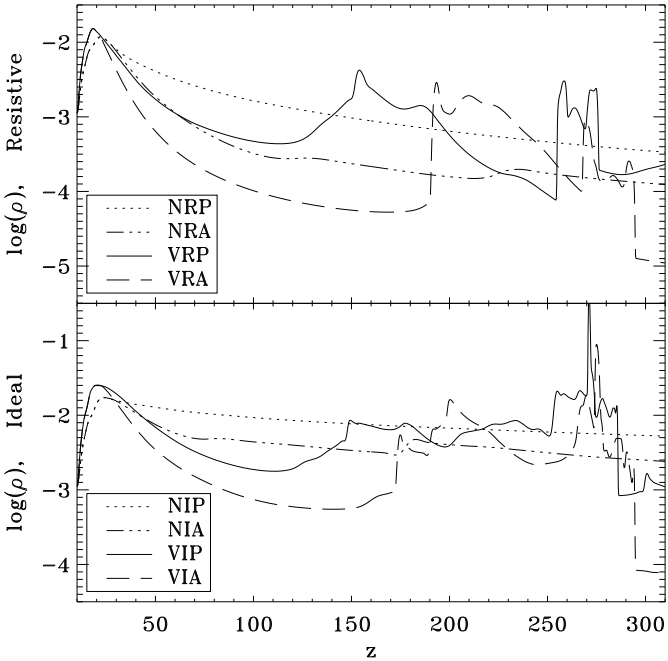}}
  \caption{Logarithmic density profiles along the axis for the ideal MHD models
    (bottom), i.e. NIP (dotted line), NIA (dashed dotted), VIP (solid), VIA
    (dashed), and the resistive MHD models (top), NRP (dotted), NRA (dashed
    dotted), VRP (solid), VRA (dashed).
    The density has on average lower values in the anti-parallel cases and also
    peaks at different locations as compared to the parallel configuration.
    Especially in the presence of resistivity, decollimation drops the densities
    by an order of magnitude and significantly amplifies the differences between
    the parallel and anti-parallel models.}
  \label{fig:vertical_profile}
\end{figure}
Fig.~\ref{fig:vertical_profile} plots the logarithm of $\rho$ along the $z$ axis
for the ideal (bottom) and resistive (top) cases.
Dotted/dashed-dotted lines are used for the non-variable parallel/anti-parallel
configurations, and solid/dashed for the variable outflows, respectively.
Density has lower overall values in the anti-parallel topology whereas flow
fluctuations alter substantially its vertical profile.
In particular, the high concentration regions are located at different heights
as compared to the parallel setup, but without demonstrating any systematic
lagging.
Both features seem to become stronger when physical resistivity is included.
On the one hand this implies the non-negligible role of numerical diffusivity
(required for stability), and on the other, the fact that larger $\eta$ values
could lead to significantly different results between the two hemispheres.
Note that since emissivity is proportional to the density squared, it is
expected that the above discrepancies could have a direct impact on
observations.
The pressure follows approximately the behavior of $\rho$ and hence it is not
shown.
As a result, the temperature has an almost flat profile for the ideal cases,
whereas Ohmic heating slightly increases the average temperature of the
resistive models, although without any distinguishable effects among them.

\begin{figure}
  \resizebox{\hsize}{!}{\includegraphics{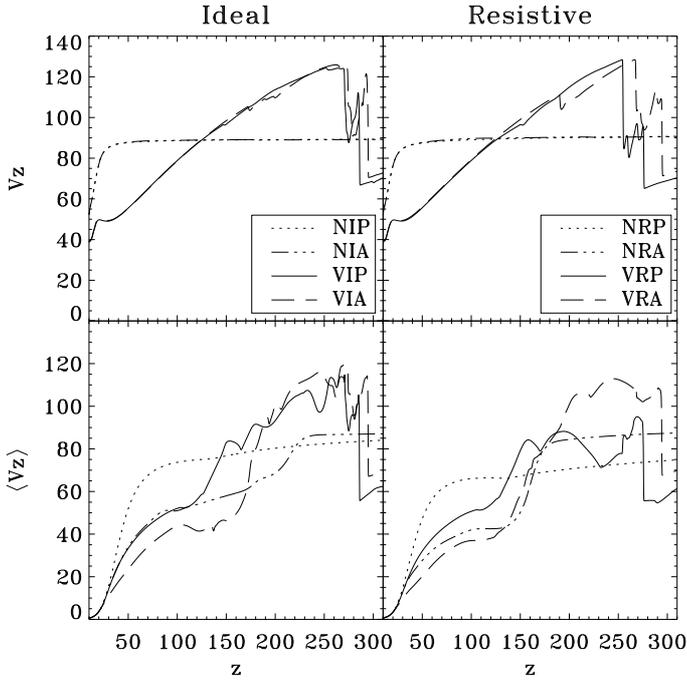}}
  \caption{Vertical velocity and weighted average of $V_z$,
    Eq.~(\ref{eq:av_vz}), along the $z$ axis.
    The line style of each model follows from Fig.~\ref{fig:vertical_profile}.
    Although the flow speed is the same among all cases, the quantity
    $\left<V_z\right>$ does demonstrate asymmetries when non-ideal MHD effects
    are included.}
  \label{fig:vertical_profile_v}
\end{figure}
In Fig.~\ref{fig:vertical_profile_v}, which follows the notation of
Fig.~\ref{fig:vertical_profile}, the vertical velocity $V_z$ (top) is plotted
versus the $z$ direction.
It was already anticipated from Fig.~\ref{fig:radial_profile} that the
parallel/anti-parallel configurations would show similar speeds.
The small displacement of the shock front between the variable resistive models
is evident here as well.
However, having found discrepancies in the radial profile of $V_z$, as well as
in the density distribution, we introduce the following weighted average flow
velocity, that incorporates emissivity in an approximate way:
\begin{equation}
  \left<V_z\right> =
    \frac{\int_0^{r_\mathrm{max}}\rho^2V_z\,2\pi r\,\mathrm{d}r}
    {\int_0^{r_\mathrm{max}}\rho^2\,2\pi r\,\mathrm{d}r}\,.
  \label{eq:av_vz}
\end{equation}
This quantity provides a better estimate for the observational implications of
the simulations than $V_z$ alone.
The integration takes place within the cylinder of radius $r_\mathrm{max} = 30$,
where $\rho$ has its highest values.
Note that for altitudes $z > 200$ the density peak along the current sheet is
not inside the integrated domain.

The bottom panels of Fig.~\ref{fig:vertical_profile_v} display the vertical
dependence of Eq.~(\ref{eq:av_vz}) for the corresponding models.
The ideal MHD regime does not show any significant discrepancies beyond
$z = 200$.
However, resistive cases do demonstrate velocity asymmetries both in the
variable and non-variable models.
The difference is of the order of $30\%$ and $20\%$, respectively, but since it
depends on the specific $\eta$ value assumed, it could be much higher or even
negligible in a real astrophysical situation.
In addition, note that including the current sheet within the integration domain
results in a slower mean speed for the anti-parallel model.
This can be seen by comparing the regions $z < 200$ and $z > 200$.
In other words, the choice of $r_\mathrm{max}$ also influences the absolute
value of the estimated discrepancy and hence the resulting percentage should not
  be taken as a robust number\footnote{On the contrary, our aim is qualitative
  and not quantitative, i.e. to show that velocity asymmetries can occur under
  some general considerations.
  A more complex average velocity could have been assumed, but we deliberately
  choose a simple function to constrain the number of unknown factors that could
  influence our results.}.

Although a parametric study of $\eta$ would clarify the asymmetry dependence on
finite conductivity, such a task is beyond the scopes of this work.
Besides, the actual resistivity value is an unknown factor in YSOs.
It suffices that even in the case of $\eta = 1.0$, a small value that does not
have significant effects on the parallel case, an asymmetry of
$\sim$$20\mathrm{-}30\%$ is found.
Nevertheless, we have performed simulations with higher and lower resistivity
values that seem to agree with the above conclusions.
We also note that the vertical flow velocity is constant along $z$ which
indicates that there is no need to simulate larger spatial scales.

\subsection{External pressure}

All simulations of pressure-confined two-component jets reach quasi steady state
configurations within a much larger simulation time as compared to the models
presented above.
This is due to the fact that the initial conditions are farther away from the
equilibrium described by the ADO and ASO solutions.
Moreover, all final configurations possess static knots along the flow axis as
well as a wave-like cylindrical outer surface that recollimates the flow at
multiple locations.
The separation of these high density regions seems to increase the lower the
ambient pressure applied.
Notably, the variability of the inner flow does not seem to disrupt this
structure, nor the shape or position of its surface.
The jet models become asymptotically cylindrical, after having passed from a
stage of oscillations in their radius, speed, density, and other physical
parameters.
Vlahakis \& Tsinganos (\cite{Vla97}) have shown that, under rather general
assumptions, this is a common behavior of magnetized outflows which start
non-cylindrically before they reach collimation.

\begin{figure}
  \resizebox{\hsize}{!}{\includegraphics{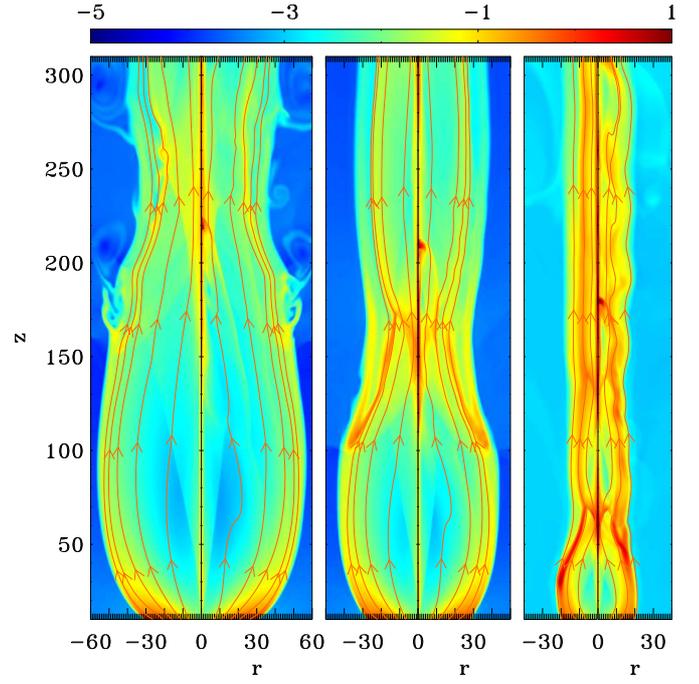}}
  \caption{Logarithmic density and field lines for models NIPL, VIPL, NIPH,
    VIPH, NIPVH and VIPVH (from left to right).
    The left and middle panels correspond to the northern and southern
    hemispheres of Fig.~\ref{fig:external_pressure}, respectively.
    The right panel displays the scenario of a very high external pressure in
    order to visualize the oscillation of the jet's cross section.
    Each panel shows the unperturbed model on the left and the time variable
    counterpart on the right.
    The outer structure as well as the condensations along the jet axis are
    static and unaffected by flow fluctuations.}
  \label{fig:NIPP-VIPP_rho}
\end{figure}
The logarithmic density along with the field lines are shown in
Fig.~\ref{fig:NIPP-VIPP_rho} for models NIPL, VIPL, NIPH, VIPH, NIPVH and VIPVH
(from left to right).
All of them have the same truncation radius, indicated in
Fig.~\ref{fig:NIP_initial}, but a different external pressure is imposed.
Despite the pressure equilibrium holding right above the bottom boundary, the
radial ram pressure of the flow expands the jet for small $z$.
Eventually it becomes compensated and then recollimation occurs, compressing the
flow at almost its initial diameter.
At higher altitudes, the jet radius increases again and the process is repeated
creating a sequence of static knots, a configuration that is stable and does not
propagate with time.
These features smooth out with $z$ and the jet acquires a cylindrical shape far
away, as seen in model NIPVH.
We have verified the stability of the static knots with simulations carried out
up to $t = 500$, though in a lower resolution.
During time evolution, Kelvin-Helmholtz instabilities appear on the interface
with the external medium, but they are gradually suppressed by the time the
quasi steady state is reached.

The jet structure of NIPH seems to be a smaller copy of NIPL, and the same holds
true for NIPVH with respect to NIPH.
This is due to the intrinsic self-similarity property of the ADO solution,
namely, all field lines have the same shape but in a different scale.
Consequently, a similar outer structure is recovered in all pressure-confined
cases, depending on the value of the external pressure imposed.
Finally note that the mixing of the two analytical outflows introduces a length
scale in the system and hence the above argument will not hold for arbitrarily
small radii.

\begin{figure}
  \resizebox{\hsize}{!}{\includegraphics{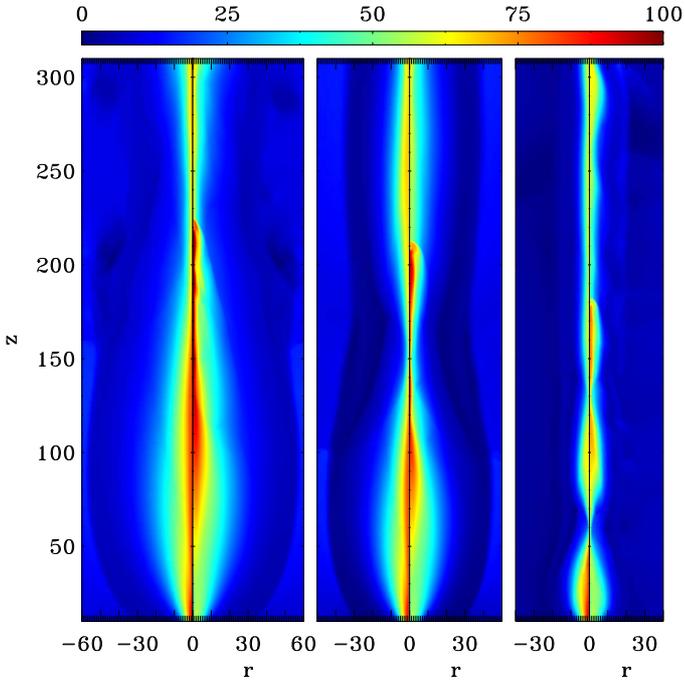}}
  \caption{Vertical velocity contours for the models shown in
    Fig.~\ref{fig:NIPP-VIPP_rho}, i.e. NIPL and VIPL (left panel), NIPH and VIPH
    (middle panel), NIPVH and VIPVH (right panel).
    The velocity maps follow the density distributions.}
  \label{fig:NIPP-VIPP_v}
\end{figure}
Fig.~\ref{fig:NIPP-VIPP_v} displays the velocity distribution for models NIPL
(left pair; left), VIPL (left pair; right), NIPH (middle pair; left), VIPH
(middle pair; right), NIPVH (right pair; left) and VIPVH (right pair; right).
As expected, the velocity profile reflects the static oscillations of the jet's
cross section since the mass flux is conserved.
In addition, its radial dependence consists of layers of decreasing speed.
Enforced time variability barely modifies the total distribution demonstrating
the stability of the configuration.
The small velocity observed in the outer medium is a product of the initial
strong transient state in which the outflow is squeezed as the truncation
surface (solid line, Fig.~\ref{fig:NIP_initial}) moves upwards.

\begin{figure}
  \resizebox{\hsize}{!}{\includegraphics{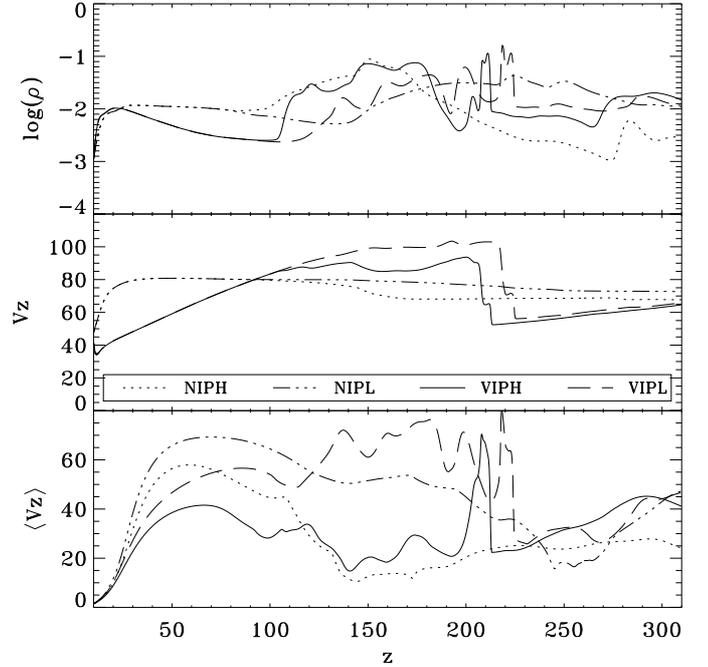}}
  \caption{The quantities $\log(\rho)$, $V_z$ and $\left<V_z\right>$
    (Eq.~\ref{eq:av_vz}) along the $z$ axis, for models NIPH (dotted), NIPL
    (dashed dotted), VIPH (solid), VIPL (dashed).
    The pressure-confined jet simulations demonstrate velocity asymmetries in
    $V_z$ and become even more evident in the calculation of
    $\left<V_z\right>$.}
  \label{fig:vertical_profile_pr}
\end{figure}
The logarithmic density, the vertical velocity and $\left<V_z\right>$ are
plotted along the axis in Fig.~\ref{fig:vertical_profile_pr} for the models NIPH
(dotted), NIPL (dashed dotted), VIPH (solid) and VIPL (dashed).
We notice that the discrepancies are stronger in the weighted velocity, a
quantity possibly relevant for observations.
Moreover, $\left<V_z\right>$ demonstrates a difference between the cases NIPH
(dotted) and VIPH (solid) which is relatively smaller as compared to the
variations between the models NIPH and NIPL.
This implies that the recollimation structure dominates, producing density peaks
much higher than those of the enforced time variability.

Velocity asymmetries are evident in this class of simulations even without
accounting emissivity.
In particular, the middle plot of Fig.~\ref{fig:vertical_profile_pr}
demonstrates a $\sim$$10\%$ deviation in $V_z$ between the two cases.
The bottom panel suggests that the observed speed of the one jet could be twice
the value of the other when weighted with the square of the density, at least
locally.
Note that in these cases the integration is calculated for
$r_\mathrm{max} = 20$.
Therefore, if a YSO outflow finds itself inside a non-uniform environment in
which the pressure ratio between the two hemispheres is as low as $2$, a
significant velocity asymmetry could be measured.

\subsection{Mass loading}

A third possible scenario is that of an increased mass loading at the base of
the outflow, such as a long lived asymmetric accretion or a strong radiation
source heating unevenly the disk surface.
In that case, the wind density can be different above and below the equator and
so will the acceleration.
However, on the one hand our numerical setup is not fully appropriate to
investigate this mechanism in a consistent way, and on the other, the results of
such simulations did not provide adequate evidence for the formation of velocity
asymmetries. 
Therefore, we briefly discuss this case without presenting an extensive
analysis.

We have carried out simulations assuming the initial setup of NIP, but setting
a two times higher density on the bottom boundary.
These models allow us to examine the readjustment of the velocity profiles
within a higher mass loading regime, naively assuming that all other jet
properties remain the same at its base.
Despite the significant modification of the boundary conditions, no considerable
asymmetries have been found.
Regarding the stellar component, we attribute this negative result to the fact
that the outflow is already super-Alfv\'enic.
A more consistent way to check how a high plasma mass will affect the velocity
is to take into account the acceleration regions.
However, this is not a straightforward task since it is not clear how the energy
input required to drive the jet will change in a high mass loading regime.
In addition, the much smaller characteristic time and lengths involved cannot be
easily included in our comparably larger scale simulations.
On the other hand, although the disk wind does include sub-Alfv\'enic regions,
they appear at large radii of low density and speed that contribute negligibly
to the velocity profile.

To sum up, the large scale two-component jet models presented in this paper,
cannot fully capture the case of an asymmetric mass loading.
The implicit assumption that the flow on the bottom boundary of each hemisphere
has different mass but exactly the same velocity might not be valid if the
acceleration mechanisms vary strongly with respect to the ejected mass.
Nevertheless, we have verified that in the cases that this holds true, no
substantial asymmetries are expected.

\section{Summary - conclusions}
  \label{sec:conclusions}

In this paper we address the velocity asymmetries of YSO outflows by
investigating two classes of candidate mechanisms that could possibly generate
this feature.

The first class depends on the intrinsic properties of the YSO and assumes a
parallel and an anti-parallel magnetic field configuration, one for each
hemisphere.
The application of physical reconnection results in different density and
velocity distributions between the two sides, leading to observable speed
discrepancies.
This scenario can provide asymmetric jets coming from intrinsic physical
conditions and processes, even without considering the complex dynamics of the
star-disk interaction.
The relative questions that arise are whether multipolar fields in the star-disk
system can exist and survive for a time scale comparable to that of jet
propagation, as well as what is the actual value of resistivity in YSO outflows,
that could amplify or even suppress the asymmetry phenomenon.

The second class of mechanisms is based on external effects, namely when the YSO
resides in a non-uniform environment.
Imposing distinct outer pressures at some boundary line of the jet is found to
directly affect the degree of collimation of each flow, which in turn results in
significantly modified propagation speeds.
This mechanism is appropriate to explain an outflow asymmetry on large time
scales.
In addition, static knots are found to manifest in the jet structure due to
multiple recollimation locations.
Their stability is robust despite the enforced flow perturbations, whereas their
separation is found to be closely associated with the ambient pressure value.
There is some evidence for stationary shocks in some systems (Matt \& B\"ohm
\cite{Ma03a}; Bonito et al. \cite{Bon11}; Schneider et al. \cite{Sch11}),
possibly associated with the collimation of the flow.
However, shocks in protostellar jets are generally observed to propagate with
the flow (Reipurth \& Bally \cite{Rei01} and references therein).
The fact that standing recollimation shocks, predicted in non-variable
simulations with a confining pressure, are not generally observed in YSO jets
implies that either the dynamical evolution of the outflow is not strongly
affected by the environment or that recollimation shocks are present but
undetected.

In the case of the RW Aur jet, it is not clear enough from the available
observations whether the asymmetry is associated with the central engine (Woitas
et al. \cite{Woi02}; Hartigan \& Hillenbrand \cite{Har09}) or with environmental
effects (Melnikov et al. \cite{Mel09}).
Both scenarios could in principle be feasible.
We have arbitrarily selected the values of our parameters to explore both
candidate mechanisms in a general manner without attempting to model this
particular object.
In order to understand the applicability of a particular model to interpret the
velocity asymmetry, more detailed observations are required, as well as more
advanced numerical models, including 3D geometry and radiative processes.

\begin{acknowledgements}
  We would like to thank an anonymous referee for helpful suggestions and the
  A\&A editor M. Walmsley for useful comments that lead to a better presentation
  of this work.
  We would also like to thank P. Tzeferacos, O. Te\c{s}ileanu and E. M. de
  Gouveia Dal Pino for fruitful discussions as well as J.-P. Chi\`eze, F. Thais,
  C. Stehl\'e and E. Audit for their support during the preparation of this
  paper.
  This research was supported by a Marie Curie European Reintegration Grant
  within the 7th European Community Framework Programme, (``TJ-CompTON'',
  PERG05-GA-2009-249164).
\end{acknowledgements}

\end{document}